\documentclass[10pt,sigconf,letterpaper,nonacm]{acmart}

\usepackage{algorithm} 
\usepackage{algpseudocode} 
\usepackage{amssymb}
\usepackage{xcolor}
\usepackage{xspace}
\usepackage{svg}

\let\emptyset\varnothing
\AtBeginDocument{%
  \providecommand\BibTeX{{%
    \normalfont B\kern-0.5em{\scshape i\kern-0.25em b}\kern-0.8em\TeX}}}

\newcommand{\edp}{\texttt{EDP}\xspace}
\newcommand{\edps}{\texttt{EDPs}\xspace}
\newcommand{\ot}{\texttt{One} \texttt{Tree}\xspace}
\newcommand{\mt}{\texttt{Multiple} \texttt{Trees}\xspace}

\begin{document}
  \pagestyle{plain}
\title[Improving the Resilience of Fast Failover Routing: \texttt{TREE}]{Improving the Resilience of Fast Failover Routing:\\ \texttt{TREE} (Tree Routing to Extend Edge disjoint paths)}

\author{Oliver Schweiger}
\affiliation{%
	\institution{University of Vienna}
		\department{Faculty of Computer Science}
	\country{Austria}
}
\email{oschweiger@cs.univie.ac.at}

\author{Klaus-Tycho Foerster}
\orcid{0000-0003-4635-4480}
\affiliation{%
	\institution{TU Dortmund}
	\country{Germany}
}
\email{klaus-tycho.foerster@tu-dortmund.de}

\author{Stefan Schmid}
\affiliation{%
	\institution{TU Berlin, University of Vienna, Fraunhofer SIT}
	\country{Germany and Austria}
}
\email{stefan.schmid@tu-berlin.de}

\renewcommand{\shortauthors}{O.\ Schweiger, K.-T.\ Foerster, and S.\ Schmid}

\begin{abstract}
Today's communication networks have stringent availability requirements and hence need to rapidly restore  connectivity after failures.
Modern networks thus implement various forms of fast reroute mechanisms in the data plane, to bridge the gap to slow global control plane convergence.
State-of-the-art fast reroute commonly relies on disjoint route structures, to offer multiple independent paths to the destination.

We  propose to leverage the network's path diversity to extend edge disjoint path mechanisms to tree routing, in order to improve the performance of fast rerouting.
We present two such tree-mechanisms 
in detail and show that they boost resilience by up to 12\% and 25\% respectively on real-world, synthetic, and data center topologies, while still retaining good path length~qualities. 
\end{abstract}

\ccsdesc[500]{Computer systems organization~Reliability}

\keywords{networks, routing schemes, fast failover, fast reroute, network resilience, data plane}

\maketitle

\section{Introduction} \label{sec:introduction}

Communication networks form a critical backbone of the digital society.
To meet their high availability and dependability requirements, these networks need to be able to deal with failures: especially link failures are unavoidable today, and are likely to become more frequent at increasing scale.
Fast reroute~\cite{DBLP:journals/comsur/ChiesaKRRS21} is an attractive solution used in most modern communication  networks today to quickly reroute traffic around failed links (henceforth also called fast failover). 
The fast reroute mechanism is implemented in the dataplane, relying only on local information for fast decision making, and hence  avoiding the overheads and delays usually involved in control plane mechanisms (such as path reconvergence, link reversal or the notification of a centralized controller)~\cite{DBLP:journals/comsur/ChiesaKRRS21}. 
Designing fast reroute mechanisms however is challenging, due to the limited information such local solutions can have when deciding to which port to forward a packet.
Indeed, even preserving connectivity is challenging~\cite{DBLP:conf/apocs/FoersterHPST21,DBLP:conf/podc/FeigenbaumGPSSS12,DBLP:journals/ton/ChiesaNMGMSS17}, and accounting for additional properties such as stretch or congestion only renders the problem more difficult~\cite{DBLP:conf/infocom/FoersterP0T19,DBLP:journals/corr/abs-2009-01497}.
A powerful and widely-used approach to realize fast reroute mechanisms is to rely on (directed) edge-disjoint spanning trees %
and paths~\cite{DBLP:journals/ton/ChiesaNMGMSS17,DBLP:journals/ton/ElhouraniGR16,DBLP:conf/infocom/FoersterP0T19,DBLP:journals/corr/ChiesaNPGSS14,DBLP:conf/infocom/FoersterKP0T21,DBLP:conf/icalp/ChiesaGMMNSS16,DBLP:conf/dsn/FoersterKP0T19,DBLP:conf/srds/FoersterKP0T19,DBLP:journals/ccr/FoersterPST18}:
upon encountering a failure, the packet can be locally steered to the next tree or path, upon which it then reaches the destination; respectively switches again after hitting the next failure.
Whereas trees use less routing table space, they are inherently limited by the global network connectivity. Edge-disjoint paths  on the other hand can directly provide routes between source and destination, hence providing better path lengths and resilience to failures.

This paper is motivated by the recent success of such disjoint path approaches, and especially \texttt{CASA}~\cite{DBLP:conf/infocom/FoersterP0T19}.
In particular, we envision that the resilience and path length quality of such edge-disjoint paths can be improved upon by attaching further subpaths, which can then act as local detours to better leverage the network's path diversity. 
We hence leverage Tree Routing to Extend Edge disjoint paths, denoted as \texttt{TREE}, and positively evaluate our new mechanism on real-world and synthetic topologies. 
To this end, we make the following contributions in this paper.
%
In this work, we: 
\begin{itemize}
    \item build upon and expand the idea of edge disjoint path structures for fast failover routing.
    \item present novel \texttt{TREE} algorithms that utilize previously unused link diversity in the network to create tree-structures to improve the overall network resilience, 
    while still 
    preserving the routing mechanism's locality.
    \item provide multiple implementations of this idea and analyze them with respect to resilience and path length and study their respective trade-offs. 
    \item conduct a comprehensive evaluation of these algorithms, both on synthetic and real data-sets. Our results show that they boost resilience by up to 12.7\% and 25.5\% respectively, retaining good path length~qualities with acceptable pre-computation overhead. 
    \item offer open access to the entire code base. 

\end{itemize}



\section{Motivation} \label{sec:motivation} 

\begin{figure}[h]
  \hspace*{-0.0cm} 
  \vspace{-3mm}
  \centering
  \includegraphics[width=1.0\linewidth]{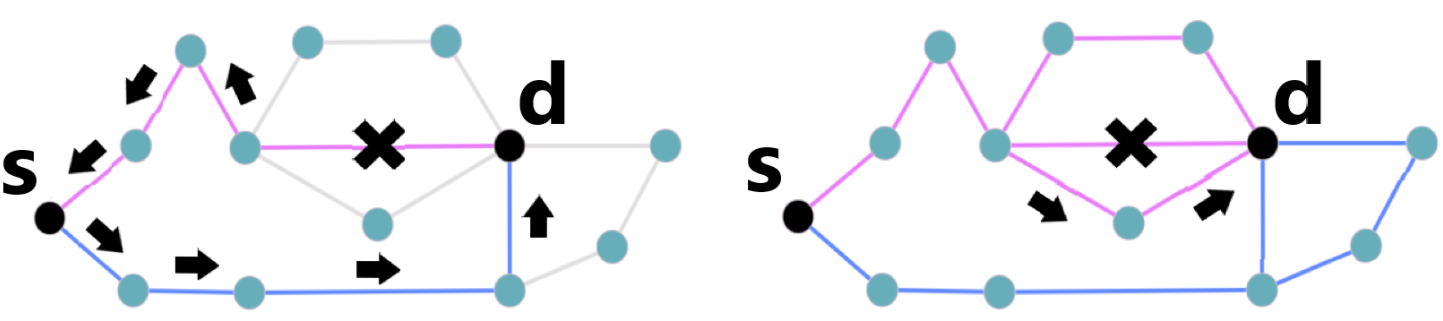}
  \caption{Motivational failover scenario. Left: Edge Disjoint Paths that require lengthy rerouting. Right: Edge Disjoint Paths extended to trees that offer shorter failover routes and improved resilience. }
  \label{fig:motivation}
  \vspace{-4mm}
\end{figure}

Prior work has shown Edge Disjoint Paths (abbreviated as \edps in the following) to offer both good resilience and path lengths for fast failover routing~\cite{DBLP:conf/infocom/FoersterP0T19}.
These \edps are precomputed for every source and destination pair and are then used to route efficiently between them. 
If a link on any path fails, packets bounce back to the source and another path is used for routing, until the destination is reached. 
However, merely using these \edps leaves a lot of resilience and path length potential on the table, as the amount of \edps that can be built is limited by the minimum cut between source and destination.
For example in the network in Fig.~\ref{fig:motivation}, only two \edps can be constructed between $s$ and $d$, leaving most of the links around $d$ unused.

We propose to enhance the \edps by extending them into trees.
By having the tree's leaves adjacent to $d$, routing along those trees in a \textit{depth first search} (DFS) manner will offer new paths to $d$ that are otherwise left unused by the \edps.
Ideally, these trees then offer fast local failover reroutes that do not require a packet to be routed back to the source, but rather jump into another branch of the tree that leads to $d$ on a short route.
Fig.~\ref{fig:motivation} illustrates this scenario: when using just \edps on the left, the link failure adjacent to $d$ leads to a lengthy rerouting process.
On the right side, when using trees, the packet can be rerouted locally and reach the destination $d$ quickly.
This can also improve network resilience: if the vertical link in Fig.~\ref{fig:motivation} between $d$ and the node beneath it fails as well, the \edps are disconnected from the destination $d$, whereas the trees would offer multiple failover routes to bridge these gaps.
We note that until convergence of routing rules, packets in this context might not be routed along shortest paths, and we discuss in parallel work~\cite{shortcut} how to \emph{shortcut} such failover routes.

\section{Model} \label{sec:model}

The network is represented as a connected graph $G = (V,E)$, with $|V| = n$ nodes connected by $|E|$ links.
These links are full-duplex symmetric, meaning they are always bidirectional. 
%
We call such a graph $k$-connected, if it remains connected after the removal of any $k-1$ links.
%

%
We will investigate fast failover routing between source nodes $s$ and destination nodes $d$, with $s \neq d$.
To this end, we assume that the static failover rules are computed ahead of time and deployed at the nodes without a priori knowledge of the link failures.
More precisely, the routing rules at a node $v\in V$ may only match on the source $s$, the destination $d$, the incident failures at $v$, and the incoming port.
Hence, the failover rules are purely local and come into effect immediately.
We do not allow for randomization, packet header modification, dynamic routing table changes, or re-convergence by means of control or data plane communication.


Regarding performance metrics, we consider two major aspects to determine overall routing quality.
First, the overall \emph{hop-count} it takes for a packet to reach $d$ from $s$. 

Second, the \emph{resilience} denotes overall routing success rate in terms of whether, in the case of link failures, $d$ could be reached at all via a given routing scheme. 
Ideally, a routing scheme provides high resilience with low hop count.


\section{Algorithm} \label{sec:algorithm}
This section describes our two methods to extend \edps to trees, in such a way that the inherent good resilience of \edps~\cite{DBLP:conf/infocom/FoersterP0T19} is maintained.

Two major variants arise, denoted as \ot (\S\ref{sec:algorithm-onetree}) and \mt (\S\ref{sec:algorithm-multipletrees}). 
\ot retains the hop count quality of \edps, whereas \mt trade in extra hop count for additional~resilience.

\subsection{\ot} \label{sec:algorithm-onetree}
The idea of \ot revolves around creating a single expansive tree out of \textit{one} of the \edps, which acts a fail-safe in case all prior paths visited on the \edps suffer from link failures.
In this version, the longest path in the \edps is being extended as outlined in Algorithm~\ref{alg:onetree}.
After this, we end up with a tree spanning potentially the whole network.
Since not every branch of this tree will lead to the destination $d$, this might lead to unnecessary detours when routing. 
Thus, we afterwards remove such unwanted branches from the tree in linear time, as shown in Algorithm~\ref{alg:truncation}.

The final result is a tree that contains no redundant paths with each leaf neighboring the destination. 
In the case of link failures, the tree's branches can act as a bridge to $d$ where traditional \edps would fail, and hence is expected to improve the overall network resilience, since it provides additional routes to $d$.
The hop count quality of \edps can be maintained in this structure, as we will discuss in \S\ref{subsec:routing}.

\begin{algorithm}[t]
	\caption{Extend - One Tree} 
	\begin{algorithmic}[1]
        \State $pathToExtend \gets$ longest path in EDPs as node array
		\For {$i=1 \to length(pathToExtend) - 1 $} \hfill{\textsc{//walk down edp}}
		    \State $nodes \gets pathToExtend$ \hfill{//\textsc{initialize new nodes as nodes of edp}}
			\State $it \gets 0$
			\While {$it < length(nodes)$}
			    \State $neighbors \gets$ neighboring nodes of $nodes[i]$
			    \For {$j=0 \to length(neighbors) - 1 $}
			        \If{$neighbors[j]$ not part of tree already}
			            \State $nodes.insert(neighbors[j])$
			        	\State add $neighbors[j]$ to tree
			        \EndIf
			    \EndFor
			    \State $it++$
			\EndWhile
		\EndFor
	\end{algorithmic} 
	\label{alg:onetree}
\end{algorithm}

\subsection{\mt} \label{sec:algorithm-multipletrees}
While \ot attempts to construct only a single tree that acts as a fail-safe attached to one the the \edps, there might still remain links left unutilized in the network. 
This is because a single tree has limitations to its growth, as shown in Algorithm \ref{alg:onetree}: newly created branches cannot cross over existing ones, as this would violate the tree's property of being acyclic. However, this could be alleviated by employing multiple trees that all act as separate structures. 

This idea lends itself especially well to our setup, as there are (most often) multiple \edps to work with. 
Thus, we propose the construction of multiple trees on top of the \edps as outlined in Algorithm \ref{alg:multipletrees}.
When examining the differences between Algorithm~\ref{alg:onetree} and \ref{alg:multipletrees},  the former one builds its routing structure on a \textit{node}-basis, whereas the latter one uses edge markings to form its trees in the network. 
This differentiation is needed, as in Algorithm \ref{alg:onetree} nodes can only belong to one single structure, whereas Algorithm \ref{alg:multipletrees} only requires the \textit{edges} of its trees to be disjoint. 
Note also how the destination $d$ is not a part of any of these trees, as the tree leaves are placed at the nodes that are \textit{neighbors} to $d$. 
As for Algorithm \ref{alg:onetree}, one needs to perform afterwards the tree truncation to every constructed tree to remove redundant branches (shown in Algorithm~\ref{alg:truncation}).

\begin{algorithm}[t]
	\caption{Extend - Multiple Trees} 
	\begin{algorithmic}[1]
        \State $EDPs \gets$ get all EDPs between $s$ and $d$, sorted desc. by their length
        \For {$i=0 \to length(EDPs) - 1 $}
		    \State $pathToExtend \gets$ $EDPs[i]$ 
    		\For {$j=1 \to length(pathToExtend) - 1 $} \hfill{//\textsc{walk down edp}}
    		    \State $nodes \gets pathToExtend$ \hfill{//\textsc{initialize new nodes as nodes in current edp}}
    			\State $it \gets 0$
    			\While {$it < length(nodes)$}
    			    \State $neighbors \gets$ neighboring nodes of $nodes[j]$
    			    \For {$k=0 \to length(neighbors) - 1 $}
    			        \If{edge to $neighbors[k]$ not part of any tree already}
    			            \State $nodes.insert(neighbors[k])$
	                        \State add edge to $neighbors[k]$ to tree

    			        \EndIf
    			    \EndFor
    			    \State $it++$
    			\EndWhile
    		\EndFor
		\EndFor
	\end{algorithmic} 
	\label{alg:multipletrees}
\end{algorithm}

\begin{algorithm}
	\caption{Truncation} 
	\begin{algorithmic}[1]
        \State $DFS \gets$ depth first search order of nodes in tree, rooted at $s$
        \State $visited \gets \emptyset$ \hfill{//\textsc{visited nodes in dfs, initialized as empty set}}
        \State $goodBranchNodes \gets \emptyset$ \hfill{//\textsc{set to mark nodes as belonging to a tree-branch to keep, initialized as empty set}}
        \State $deleteMode \gets$ false \hfill{//\textsc{flag that determines wether traveled branches are do be deleted}}
		\For {$i=0 \to length(DFS) - 1 $} \hfill{//\textsc{perform DFS}}
		    \If{going forward in DFS}
	            \State visited.insert(DFS[i]) \hfill{//\textsc{mark node as visited}}
            \Else
                \State visited.remove(DFS[i]) \hfill{//\textsc{remove node from visited}}
            \EndIf
            \If{going forward in DFS $\And$ $goodBranchNodes$ contains $DFS[i]$}
                \State $deleteMode \gets$ false
            \EndIf
            \If{$deleteMode$}
                \State remove $DFS[i]$ from tree
            \EndIf
            \If{$DFS[i]$ is a leaf $\And$ $DFS[i]$ not neighboring $d$}
                \State $deleteMode \gets$ true
            \ElsIf{$DFS[i]$ neighboring $d$}
                \State goodBranchNodes.add(visited) \hfill{//m\textsc{ark all visited nodes as a good branch}}
            \EndIf
		\EndFor
	\end{algorithmic} 
	\label{alg:truncation}
\end{algorithm}

\subsection{Routing}\label{subsec:routing}
Now that we have established how both structures are computed, the question remains how to efficiently route along them. Summing up, the following steps are to be carried out: 
\begin{enumerate}
    \item compute the set of \edps from $s$ to $d$ 
    \item sort the set of \edps in descending order\footnote{By doing this, we start applying the tree formation to the longest \edp first. This is useful, as the shortest \edps tend to remain in their original form and thus offer the fastest route to $d$ in case they do not suffer from any failure.}
    \item apply either Algorithm \ref{alg:onetree} or \ref{alg:multipletrees} to the \edps
    \item truncate the trees to remove unwanted branches
    \item sort the original set of \edps in ascending order
    \item route from $s$ to $d$ along the \edps one after another 
    in a \textit{depth-first} order (henceforth called \emph{Depth First Routing} (\emph{DFR})), until $d$ is reached. By the previous sort, we now route to $d$ by trying out the shortest paths first. 
\end{enumerate}

Notice how depth-first traversal of the \edps also makes sense for the \edps left untouched, as the normal routing-behaviour of going down and up (in case of failure) a single path is equivalent to a depth-first traversal.

There is one additional trick that we apply to DFR. While traversing the tree depth-first, at nodes with a degree greater than 2, we make use of an internal per-node ranking that determines the order at which the individual branches of the tree are visited. This rank indicates which branch has the shortest distance to any leaf below it. This then has the effect that, when routing, the shortest potential path to $d$ within the tree is always picked.

Moreover, we extend DFR for \ot to retain the hop count quality of \edps, under the assumption that the \edps can successfully reach the destination $d$.
The single tree of \ot still contains the original \edp as one of its branches and we adapt the routing to first try this path, before using additional tree branches.
In this way, the hop count of \edps and \ot is identical, as long as the \edps still connect source and destination.

Ranked DFR can be achieved by static routing tables. 
As the \edps and tree-extensions are pre-computed, these rules need only be configured at the individual nodes, where for packets from $s$ to $d$, the outgoing port can be uniquely identified by the incoming port and incident failures.

\medskip


\section{Evaluation} \label{sec:evaluation}

This section evaluates how both tree algorithms perform in comparison to \edps~\cite{DBLP:conf/infocom/FoersterP0T19}.
For this, we carried out a multitude of experiments, both on real and synthetic data sets and investigate two failure models that help with examining the strong and weak points of the respective techniques.
We consider two performance measures to evaluate performance: the \textbf{hop-count} $h_{rs}$ of an employed routing scheme $rs$, and its \textbf{resilience} $r_{rs}$. Both metrics have been introduced in \S\ref{sec:model}. 

\medskip

\subsection{Link Failure Models}
Our setup revolves around applying failures to the network, i.e., the removal of links. We use two failure models:
\begin{itemize}
    \item \textbf{Random}: edges in the network fail at random. \vspace{2mm}
    \item \textbf{Clustered} 
    simulates failures around $d$ with a chance of $p^{f}$. Additionally, failures propagate to adjacent edges, with a probability being reduced per hop by $p^{f}_{\Delta}$. For our experiments, we set $p^{f}_{\Delta}=0.3$.
    This model aims to simulate the failure behavior of zones affected by some kind of natural disaster~\cite{DBLP:series/ccn/RH2020,DBLP:conf/infocom/TapolcaiRVG17}, such as earth quakes or other sources of extensive power outages, where network links might fail in an area of larger scale around an, e.g., epicenter~\cite{DBLP:journals/ton/NeumayerZCM11,DBLP:journals/ton/AgarwalEGHSZ13}.
\end{itemize}

\medskip

Note: the meaning of \textit{failure rate} differs between clustered and random failures. Clustered failures describe a certain percentage of failing links around $d$ (which then propagate further). For random failures, we treat the failure rate proportional to the general network connectivity (described by $k$), as to better generalize this metric for both dense and sparse graphs. As such, a failure rate $p^{f}$ for k-connected graphs means that $p^{f}*k$ links fail, chosen uniformly at random.

\subsection{Evaluation Setup and Metrics}
We examine both real-world topologies by routing on graphs from the Internet Topology Zoo~\cite{DBLP:journals/jsac/KnightNFBR11}, as well as artificial random Erdős-Rényi graphs~\cite{gilbert1959random,Erdos:1959:pmd} and data center topologies~\cite{DBLP:conf/nsdi/SinglaHPG12}.
The construction of these Erdős-Rényi graphs is defined by two essential parameters $n$ denoting the amount of nodes in the graph, and $p$ denoting the probability of an edge appearing between two nodes.

For generating meaningful measurements of $h_{rs}$ and $r_{rs}$, we simulated 1000 routing runs per data point in our plots in the next Sections~\ref{subsec: randomGraphs} and~\ref{subsec:topozoo-eval}.
Each run, for random graphs, a new Erdős-Rényi graph is generated with the same parameters $n$ and $p$, and also $s$ and $d$ are newly chosen at random for every run. 
For the real-world topologies, we keep the same graph over all runs, but also pick $s$ and $d$ randomly---as well as for the data center topologies, with 200 instead of 1000 runs each.
%

\medskip

We compare our \ot and \mt with the state-of-the-art fast failover routing scheme for \edps~\cite{DBLP:conf/infocom/FoersterP0T19}, which we implement in Python 3.8.6 using NetworkX 2.5~\cite{SciPyProceedings_11}. 

%

For the three schemes, we compute and compare the average resilience and hop count for given failure rates in \S\ref{subsec: randomGraphs} and \S\ref{subsec:topozoo-eval}, as well as the average pre-computation time in \S\ref{sec:computation}.
For comparing schemes in terms of their hop count, we only take into account the runs where the \edps succeeded.\footnote{By construction, when \edps reach the destination, both \ot and \mt do so as well, as they extend the \edps.}

%
%

\newpage

\subsection{Routing in Random Graphs} \label{subsec: randomGraphs}
\subsubsection{Random Failures}

\begin{figure}[t]
  \hspace*{-0cm} 
  \centering
  \includegraphics[width=1\linewidth]{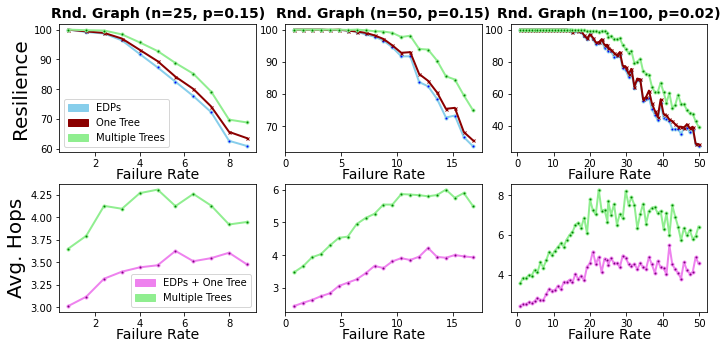}
  \caption{Random graphs with random failures.}
  \label{fig:random_random}
\end{figure}

Regarding random failures in the network, Figure \ref{fig:random_random} plots the average results for three different random graph sizes, the parameters $n$ and $p$ of which can be found in their respective label.
The top row shows that 
\mt outperforms \ot and \edps in terms of resilience, offering resilience gains of up to 21.5 percent, while \ot also outperforms the \edps, though less drastically.
The resilience gains of \ot of up to 5 percent is still noteworthy, since it offers the exact same average hop count as the \edps (for this reason, their \textcolor{blue}{blue} and \textcolor{red}{red} lines from the top row are joined into a single \textcolor{violet}{violet} one for the bottom row in Figures \ref{fig:random_clustered} and \ref{fig:zoo_random}).
On average, it takes \mt 1.51 hops longer to reach $d$, compared to \edps or \ot.

\subsubsection{Clustered Failures}
\begin{figure}[t]
\vspace{-8mm}
  \hspace*{-0cm} 
  \centering
	\includegraphics[width=1\linewidth]{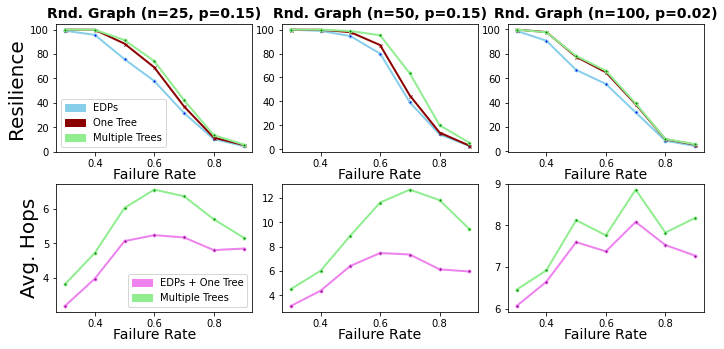}
  \caption{Random graphs with clustered failures.}
  \label{fig:random_clustered}
	\vspace{-2mm}
\end{figure}

Similar results (shown in Fig. \ref{fig:random_clustered}) arise when comparing routing performance on the random graphs generated with the same parameters $n$ and $p$ as before, but now with clustered failures. 
\ot outperforms the \edps in terms of resilience by 12.7 percent, while offering the same average hop count. 
\mt manages to improve resilience by up to 24 percent, while still offering a very comparable hop count (an average increase of 1.6 hops), especially for larger, less dense graphs (i.e., $n=100$, $p=0.02$). 

\subsection{Routing in Real-World Topologies}\label{subsec:topozoo-eval}
For testing our algorithms on real-world topologies, we opted to use graphs from the Internet Topology Zoo~\cite{DBLP:journals/jsac/KnightNFBR11}, as common in related work.
Overall, we chose three different topologies with varying $n$, in order to evaluate performance on both larger and smaller real-world networks.

As with the artificial graphs from \S\ref{subsec: randomGraphs}, we apply both random and clustered failures to the graph, while pushing up the failure rate, and compare resilience and average hop count of the three schemes.

\subsubsection{Random Failures}

\begin{figure}[t]
  \hspace*{-0.0cm} 
  \centering
  \includegraphics[width=1\linewidth]{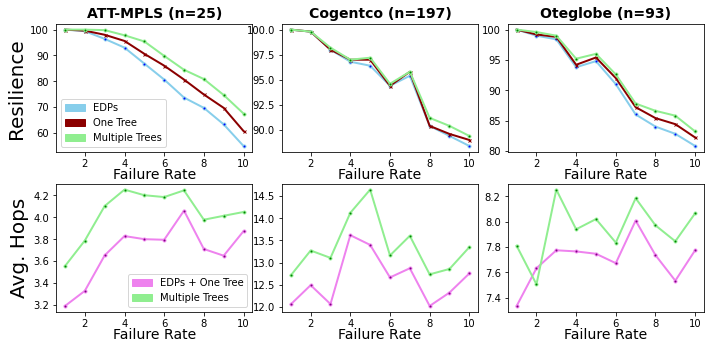}
  \caption{Real-world graphs with random failures.}
  \label{fig:zoo_random}
	\vspace{-4mm}
\end{figure}

Fig.~\ref{fig:zoo_random} shows that, similarly to the artificial networks, both \ot (resilience boost up to 7 percent) and \mt (resilience boost up to 12.6 percent) outperform the \edps on real-world graph for random failures, \ot often times offering resilience values close to those of \mt. 
\ot and \edps take the same amount of hops to reach $d$, with \mt exhibiting an average increase in hop count by 0.44 hops.

\subsubsection{Clustered Failures}

\begin{figure}[t]
  \hspace*{-0.0cm} 
  \centering
  \includegraphics[width=1\linewidth]{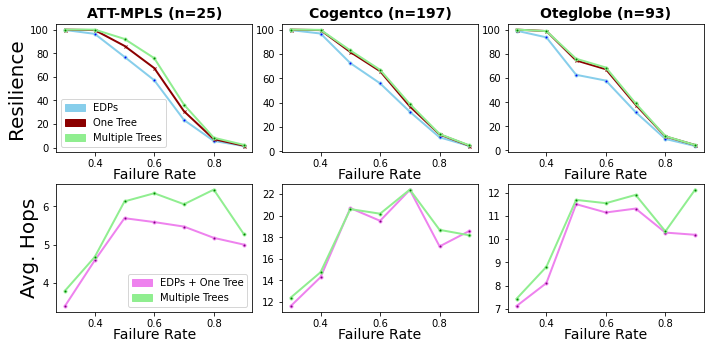}

  \caption{Real-world graphs with clustered failures.}
  \label{fig:zoo_clustered}
	\vspace{-4mm}
\end{figure}

Next, we employ clustered failures to the real-world topologies. Yet again, a similar pattern emerges as shown in Figure \ref{fig:zoo_clustered}.
\ot and \mt boost resilience by up to 11.9 and 18.7 percent respectively, with \ot being just as fast as the \edps. \mt also shows to be efficient here, as it offers high resilience improvements while only increasing the hop count by 0.52 on average.

\subsection{Routing in Data Center Topologies}\label{sec:datacenters}
Recent research~\cite{DBLP:conf/nsdi/SinglaHPG12,DBLP:conf/sigcomm/KassingVSSS17,DBLP:journals/algorithmica/DinitzSV17,DBLP:conf/conext/ValadarskySDS16,DBLP:conf/hotnets/ValadarskyDS15,DBLP:conf/esa/DinitzSV15} has shown significant performance benefits for flat topology designs in data center networks and we as hence also investigate our tree algorithms in these settings.
For easy scalability of our evaluations, we choose topologies as in Jellyfish~\cite{DBLP:conf/nsdi/SinglaHPG12}, modeled as random $\delta$-regular graphs.
Multiple test runs with $\delta=8$ and $n \in \{25,50,100\}$ were performed, again both with clustered and random failures. For \mt, in these experiments resilience could be improved by up to 25.5\% at an average hop increase of 4.5 for $n=50$ for clustered failures, and a resilience-increase of up to 18.5\% at an average hop increase of 1.96 for $n=25$ for random failures. However, \ot performed slightly worse on these graphs, as performance was only marginally better than its \edp counterpart most of the time.
Notwithstanding, in general, the performance of \mt and \ot is similar to the settings in \S\ref{subsec: randomGraphs} and \S\ref{subsec:topozoo-eval}.

\subsection{Computational Runtime Comparison}\label{sec:computation}
\begin{figure}[t]
  \centering
  \includegraphics[width=0.9\linewidth]{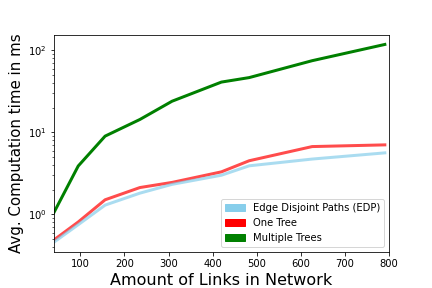}
  \caption{The relationship between computation time in milliseconds and the amount links in the network for the three compared algorithms }
  \label{fig:runtime}
\end{figure}

We next analyze the runtime that it takes for the \edps, \ot, and \mt structures to be computed.
Our single-threaded computations were performed on a Ryzen 9 3900X CPU @3.8GHz with 32GB of DDR4 RAM, using Python 3.8.6 with NetworkX 2.5~\cite{SciPyProceedings_11}.
To this end, we use Erdős-Rényi graphs with $n$ rising from 25 to 105 in steps of 10, a constant $p=0.15$ and taking 200 runs for each $n$.
We plot in Fig.~\ref{fig:runtime} the runtime of the three algorithms w.r.t.\ to the number of links in the network, taking the average runtime for ranges of size 5, i.e., 1 to 5 links, 6 to 10 links, etc.

In comparison to the \ot and \mt structures being computed, we observe a much heavier workload for the creation and truncation of \mt.
We see that the computation times for \mt scales linearly 
with the number of links multiplied with the number of trees built in the network, while \ot computes much faster, even in larger networks, as there is only a single tree to build and, consequently, truncate: there is not much overhead when comparing \ot to \edps.

\subsection{Discussion}
Overall, these results showcase the unused potential in the random and real-world graphs that our techniques leverage.
By building trees from the \edps, we manage to establish local failover mechanism that increase resilience. 
\ot retains the path length of \edps, while still providing noticeable resilience gains.
The two rows in our plots from Fig.~\ref{fig:random_random} to \ref{fig:zoo_clustered} also exemplify the trade-off that emerges between the \ot and \mt techniques: 
\ot keeps the efficiency of the \edps in terms of hop count, while still leaving some potential links unused.
Consequently, \mt does use these unused links to further increase resilience, with the downside of exhibiting (most often not drastically) larger hop-counts. 
Thus, we perceive the use of either \ot or \mt to be highly dependable of the exact use-case and underlying network properties. 
If the speed of packet arrival is not a top priority, but a higher resilience is desired, one might use \mt. 
Otherwise, if one might want to keep hop-counts short but still leverage some unused link-potential and computation time for the routing structures is critical, we propose to construct \ot on top of the existing \edps.
Lastly, however a downside of our new algorithms is the additional routing table space consumption, and it would be interesting to adapt our algorithms to find a good trade-off respectively to use compression~\cite{DBLP:conf/hotnets/StephensCR13,DBLP:conf/sosr/StephensCR16}.


\section{Related Work} \label{sec:relatedwork}
Failures are common in computer networks~\cite{DBLP:series/ccn/RH2020} and hence fast reroute mechanisms have been studied comprehensively over the last decades, we refer to a recent survey for an overview~\cite{DBLP:journals/comsur/ChiesaKRRS21}.
In general, for rapid connectivity restoration, the computation \& distribution of new routes is too slow~\cite{DBLP:conf/nsdi/LiuPSGSS13}, but even convergence mechanisms in the data plane such as \texttt{DDC}~\cite{DBLP:conf/nsdi/LiuPSGSS13} can introduce non-trivial delays~\cite{DBLP:conf/spaa/BuschST03}.

As such many lines of research have considered how to overcome the delay induced by convergence mechanisms by means of pre-computation; our work falls into this regime.

Some of these fast failover algorithms leverage packet header modification, e.g., to carry failure information~\cite{DBLP:conf/sigcomm/LakshminarayananCRASS07,DBLP:conf/infocom/FoersterPCS18}, by employing MPLS~\cite{mplsfrr} or to leverage graph exploration approaches~\cite{hotsdn14failover}, but they can be disruptive to other network functions and moreover require specialized equipment or in-node computation abilities, similar arguments can be made for mechanisms that require dynamic state on the forwarding devices, such as, e.g., resilience via a rotor router~\cite{parallel-rotor}.
Recent work also investigated the use of randomization~\cite{icalp16,icnp19}, but they require some means of random number generation on the routers and can lead to packet reordering.

In contrast, many deterministic static fast failover algorithms, as in our work, do not suffer from the above issues, however at the price of reduced resilience.
Conceptually, in this framework, achieving perfect resilience is impossible~\cite{DBLP:conf/podc/FeigenbaumGPSSS12,DBLP:conf/apocs/FoersterHPST21}, i.e., to guarantee reaching the destination under the assumption of connectivity.
State-of-the-art deterministic fast failover algorithms hence leverage underlying pre-failure connectivity assumptions to pre-compute disjoint routing structures, between which the packets can be locally migrated after hitting a failure.
For destination-based routing, the standard approach is by arc-disjoint rooted spanning trees (arborescences)~\cite{DBLP:journals/corr/ChiesaNPGSS14,DBLP:journals/ton/ElhouraniGR16,DBLP:journals/ton/ChiesaNMGMSS17}, yielding worst-case resilience guarantees related to the number of such trees that can be constructed. 
Recent work expanded upon their construction in heterogeneous networks in a heuristic fashion, by, e.g., adding local non-spanning arborescences~\cite{DBLP:conf/infocom/FoersterKP0T21}, or by employing partial structural networks~\cite{keep-fwd}.
Nonetheless, due to the paradigm of destination-based routing, their path length falls behind source-destination-based approaches~\cite{DBLP:conf/infocom/FoersterP0T19}.

When routing based on both source and destination, \texttt{CASA} showed that the use of edge disjoint paths (\edps) can improve worst-case resilience guarantees and improve the hop count, at the expense of additional routing table space.
Our work is most related to \texttt{CASA} and builds upon it, by extending their approach of \edps with one or multiple trees.
Our \ot approach retains the hop count quality of \texttt{CASA}, but adds additional resilience, whereas our \mt approach shows longer path lengths in our evaluation, but improves even upon the resilience of \ot.

\section{Conclusion} \label{sec:conclusion}
We introduced a method that builds upon established static and deterministic local fast failover routing rules and further modifies and enhances them.
Two algorithms were introduced: \ot manages to keep the hop count low as in traditional approaches using Edge Disjoints Paths, but boosts overall network resilience.
\mt further enhances the network resilience by extending all \edps to trees, at the cost of slightly larger hop counts.
We analyzed these trade-offs in evaluations on real-world, random, and data center topologies and saw good performance benefits over previous work, with, e.g., \mt providing up to over 25 percentage points of additional routing success.

\subsection{Outlook}
We believe that there are multiple promising research directions that go beyond the scope of this short paper.
First, it would be interesting to consider different approaches to tree generation and traversal, and also to investigate the idea of extending not just one but also not all \edps to trees, e.g., by capping the number of trees by some parameter.
Second, additional performance measures could be investigated by, e.g., incorporating the use of balanced incomplete block designs for several $sd$-flows as in \texttt{CASA}~\cite{DBLP:conf/infocom/FoersterP0T19} to study the congestion effects of fast reroute mechanisms.
Additional venues for future research could be the use of randomized algorithms, but also the analysis and extension of our algorithms for further topologies and failure models, e.g., for data centers and shared risk link groups

\vspace{1mm}
\noindent\textbf{Reproducibility.} In order to foster reproducibility and to enable comparisons with future work, our source code will be made available at \url{https://github.com/oliver306/TREE}.

\vspace{1mm}
\noindent\textbf{Acknowledgments and Bibliographical Note.}
This paper appears~\cite{treeancs} at the ACM/IEEE Symposium on Architectures for Networking and Communications Systems 2021 (ANCS'21) and we thank our shepherd Brent Stephens and the reviewers for their valuable feedback. 
We also thank Paula-Elena Gheorghe for an earlier implementation of a tree algorithm.
This research received funding from the Vienna Science and Technology Fund (Wiener Wissenschafts-, Forschungs- und Technologiefonds WWTF) project, Fast and Quantitative What-if Analysis for Dependable Communication Networks (WHATIF), ICT19-045, 2020-2024.

\balance
\bibliographystyle{ACM-Reference-Format}
\bibliography{acmart}
\end{document}